# La Loi organique relative aux lois de finances (LOLF) dans les institutions culturelles publiques du spectacle vivant en France

**Ammar KESSAB**
Doctorant en Sciences de Gestion
UMR-MA 49 GRANEM, Université d'Angers.
UFR de Droit, Economie et Gestion
13, Allée François Mitterrand, 49036 Angers Cedex 01, France.
**ammar.kessab@univ-angers.fr**

**Résumé**

Dans un contexte de crise des finances publiques, la France fonde aujourd'hui tous ses espoirs sur l'« évaluation de la performance » pour amortir les effets d'une crise complexe. Sous le signe de la « modernisation de l'Etat », une nouvelle « constitution financière » appelée Loi organique relative aux lois de finances (LOLF) est devenue le levier principal des réformes de la gestion publique.
Pleinement appliquée sur les affaires culturelles depuis 2006, la LOLF repose sur un ensemble d'indicateurs de performance et fixe aux institutions culturelles publiques du spectacle vivant des objectifs précis à réaliser. Cet article définit et analyse le schéma de la conception et du parcours de ces indicateurs ainsi que les cibles-objectifs à atteindre. Il examine aussi la polémique générée par ce nouveau mode de gouvernance.

**Mots-clés**
**LOLF, évaluation de la performance, indicateurs de performance, spectacle vivant, France**

**Introduction**

Les premières tentatives de réformer la loi organique du pilotage des dépenses publiques dès son application en 1959 marquent le début de la prépondérance de la question purement financière dans le processus des réformes de la gestion publique en France. Influencé par les programmes américains de rationalisation budgétaire à partir de la moitié des années 1960 [B. Perret, 2006], par la doctrine mondialisée du *New public management* dans les années 1970 [P. Bezes, 2005] et par le mouvement général du pilotage par la performance des pays de l'Organisation de coopération et de développement économiques (OCDE) au début des années 1990 [C. Chol et F. Waintrop, 2003], ce processus des réformes intégrera tour à tour la « rationalisation », l'« évaluation » puis la « performance » dans ses principes d'action. Ainsi, l'« évaluation de la performance » deviendra-t-elle le levier principal de la « modernisation de l'Etat » [A. Kessab, 2009] et se concrétisera, après l'abrogation de l'ex-intouchable loi de 1959, par l'application effective en 2006 d'une nouvelle « constitution financière » appelée Loi organique relative aux lois de finance (LOLF).

Comme dans de nombreux pays occidentaux, deux raisons essentielles ont poussé la France à chercher cet « impératif » de gestion par la performance. La première raison est le coût budgétaire apparemment excessif de la fonction publique et des administrations [J. Chevalier et L. Rouban, 2003]. La seconde tient aux citoyens qui, de plus en plus initiés à la politique et à l'économie, sont devenus plus exigeants et demandent à l'Etat de leur rendre compte *(accountability)*, et ceci dans un contexte budgétaire frêle et une situation macroéconomique dominée par la concurrence [C. Chol et F. Waintrop, 2002]. *« […] c'est dans les moments où l'argent public se fait plus rare que le besoin d'en contrôler l'usage apparaît plus que jamais indispensable voir même vital pour la pérennisation de l'organisation collective. »* [M. Bouvier, 1997].

En France, par souci de l'exception culturelle et pour d'autres considérations historiques encore, l'Etat est l'entrepreneur majeur dans le secteur culturel : il est le deuxième financeur de l'activité culturelle en général (après les collectivités territoriales) et le premier financeur du secteur public du spectacle vivant auquel il contribue à lui donner un caractère de grande dépendance aux subventions. Il est aussi le premier contrôleur des missions et des financements notifiés aux institutions culturelles publiques du spectacle vivant qui sont régies par des statuts ou des conventions les engageant à assurer une mission d'intérêt général.

Eléments stratégiques de la définition et la concrétisation de la politique culturelle nationale, les institutions culturelles publiques du spectacle vivant ne sont pas restées en dehors des grandes transformations systémiques et structurelles provoquées par l'histoire mouvementée des réformes de la gestion des finances publiques en France. En effet, ces institutions ont dû, dès 1970, appliquer des circulaires, des décrets, des chartes et autres « obligations de services » rédigés plus ou moins sous l'influence de la réforme générale de chaque époque, et à travers lesquels l'Etat évaluait les activités de ces entreprises publiques.

Intégrée en 2006 à cette panoplie de textes hérités des l'histoire des réformes, la LOLF est venue renforcer par ses indicateurs de performance une culture du résultat déjà existante dans le but de rationaliser les dépenses [C. Albanel, 2008], mais surtout pour inciter les institutions culturelles publiques du spectacle vivant à réaliser l'objectif central de la politique culturelle française : la sacro-sainte démocratisation de la culture, jugée inaccomplie depuis plus de 40 ans [J.C. Wallach, 2006 ; N. Sarkozy, 2007].

Qu'elles soient sous la tutelle directe ou indirecte du ministère de la Culture, la plupart des institutions culturelles publiques du spectacle vivant ont été soumises à la LOLF à partir de 2006.

Les Centres dramatiques nationaux (CDN), au nombre de trente trois (33), sont à l'avant-garde de l'application du nouveau schéma de performance et pour cause, c'est les seuls établissements de « production et de diffusion du spectacle vivant » qui sont financés à plus de 50% par l'Etat [J. F. Hirsch, 2009]. Ils sont en avance par rapport aux autres établissements de production et de diffusion du spectacle vivant quant à l'alignement sur les normes LOLF. Ils constituent de ce fait, avec les soixante-dix (70) Scènes nationales (SN) soumis tout autant à la LOLF mais seulement depuis 2008, le meilleur cas d'étude pour extraire et analyser le nouveau schéma de la performance. Nous appellerons de ce fait « institutions culturelles publiques du spectacle vivant » l'ensemble des CDN et des SN qui représentent 87% de l'ensemble des institutions théâtrales en France.

Les CDN sont des structures dirigées par un ou plusieurs artistes, financées en moyenne à hauteur de 60% par l'Etat avec lequel un contrat appelé « de décentralisation dramatique » les lie pour une période de trois ans. Les Scènes nationales sont des théâtres publics labélisés par le ministère de la Culture et soumis à une circulaire qui précise leurs droits et leurs responsabilités.

**La Loi organique relative aux lois de finances dans les institutions culturelles publiques du spectacle vivant**

Après un consensus politique sans précédent, la Loi organique relative aux lois de finances a été pleinement mise en application en janvier 2006. La LOLF est à la fois l'amorce d'une réforme et une révolution. Amorce d'une réforme, parce qu'elle représente un aboutissement d'un ensemble de manœuvres directes et indirectes de réformes. Révolution, en ce sens où elle a remplacé une ordonnance (loi de 1959) barycentre autour de laquelle toutes les lois de finances orbitaient.

La LOLF est une nouvelle conception de la gestion publique. Elle s'appuie sur quatre éléments fondateurs : une nouvelle structure budgétaire (triade : missions/programmes/actions) ; de nouvelles règles de gestion (fongibilité des crédits) ; une accentuation sur la transparence (indicateurs de performance) ; un plus grand pouvoir du Parlement.

Sur le plan technique, la LOLF impose aux gestionnaires de rendre des comptes sur l'efficacité de l'utilisation des crédits qui leur ont été attribués [A. Lambert et D. Migaud, 2005]. Elle instaure des Projets Annuels de Performance (PAP) qui présentent les actions des différentes administrations pour l'année à venir. Et l'évaluation des objectifs se fait l'année suivante dans les Rapports Annuels de Performance (RAP). Les différentes missions sont évaluées à travers un ensemble d'indicateurs de performance.

Comme nous l'avons signalé dans l'introduction, la performance ou la culture du résultat s'appliquait sur les institutions culturelles publiques du spectacle vivant bien avant l'instauration de la LOLF. Des indicateurs chiffrés de performance, construits pour la plupart singulièrement et selon la spécificité de chaque structure ont étaient fixés dès 1995 pour les Centres dramatiques nationaux (Contrats de décentralisation) et dès 1997 pour les Scènes nationales (Contrats d'objectifs). Ces indicateurs sont toujours en vigueur et sont exploités hors le cadre de la nouvelle loi organique. L'originalité des indicateurs de performance selon la LOLF réside dans le fait qu'ils sont standardisés et s'appliquent sur l'ensemble des institutions sans distinction. De plus, la LOLF a détabouisé une notion d'évaluation de la performance dont les agents de l'Etat au niveau central avaient des réticences pour son utilisation eu égard de la pression idéologique cultivée par des acteurs du terrain farouchement hostiles à ce qu'ils considèrent comme un vrai danger pour la création et pour leurs emplois.

Prise en charge par le ministère de la Culture et de la Communication, la mission « Culture » selon la LOLF se répartit en trois programmes ministériels et un programme interministériel. Deux programmes concernent le spectacle vivant :
• Le programme 131 « Création »
• Le programme 224 « Transmission des savoirs et démocratisation de la culture »

Chaque programme se subdivise en actions. Au programme sont associés des objectifs. L'objectif est mesuré par au moins un indicateur et au plus quatre, nommés indicateurs de performance.
80% des crédits du programme 131 sont destinés au spectacle vivant [J. F. Hirsch, 2009] dont la plus grande partie constitue les subventions annuelles que reçoivent les institutions publiques du spectacle vivant qui doivent par conséquent se mobiliser pour atteindre les objectifs fixés et renseigner chaque année les indicateurs de performance.
Par contre, le programme 224 est un programme transversal à l'ensemble des directions du ministère : les indicateurs ne portent pas tant sur des institutions que sur des dispositifs mis en place par le ministère.

**Les indicateurs de performance selon la LOLF**
Un indicateur de performance est un moyen quantitatif qui s'appuie sur un ensemble de données chiffrées dont des statistiques et des ratios qui permettent d'évaluer un objectif donné [P. Jeannin, 2008].
La littérature existante sur la question des indicateurs de la performance dans le secteur culturel [P.J. Ames, 1994 ; G. Brosio, 1994 ; P.M. Jackson, 1994 ; J.M. Schuster, 1996 ; G. Evans, 2000 ; R. Towse, 2001 ; P. Vanden Eeckaut, 2002 ; G. Pignataro, 2003 ; P. Jeannin, 2008] pose essentiellement la question suivante : mesurer quoi et comment ?
L'utilisation des indicateurs dans l'activité culturelle et artistique est née de l'impossibilité pour les financeurs privés ou publics d'exploiter des signaux émis par le marché pour évaluer les aspects de la production artistique des structures qu'ils financent [G. Pignataro, 2003]. En effet, l'activité culturelle, très rarement portée dans une dimension commerciale à but lucratif, ne peut pas réagir au marché de la même façon qu'une structure privée qui elle, doit jouer le jeu de l'offre et la de la demande, et par conséquent, déverser dans son espace d'activité des informations détectables puis exploitables. Pour remédier à cette déficience, les financeurs ont alors construit des unités de mesure « virtuelles » de performance pour juger empiriquement la valeur des productions artistiques. Cet outil est censé fournir une référence objective pour évaluer une matière symbolique difficile à estimer dans son aspect économique.
Les indicateurs de performance, quantitatifs ou qualitatifs, sont utilisés sur le plan microéconomique pour définir le rendement des structures culturelles, ou sur le plan macroéconomique, pour orienter les politiques de rationalisation.

Deux sortes d'indicateurs existent : les indicateurs appelés « de description » ou « d'observation » et les indicateurs dits « d'évaluation » ou de « nouvelle génération ». Les premiers sont utilisés pour décrire les caractéristiques de la production artistique et de la consommation et sont utilisés en France depuis 1973 dans les enquêtes sur les pratiques culturelles des Français et autres statistiques culturelles du ministère de la Culture. Le deuxième type d'indicateurs a été introduit et généralisé en France dans le cadre de la LOLF.
Concernant le support de l'évaluation, on peut distinguer aussi deux différents objets mesurables : les productions artistiques et les objectifs de ces mêmes productions. Le premier objet est facilement quantifiable du moment où les informations nécessaires pour le mesurer à travers son volume sont disponibles. Le deuxième pose, par contre, la fameuse problématique de la quantifiabilité du qualitatif artistique comme par exemple l'extrême complexité d'identifier l'impact émotionnel et intellectuel des activités artistiques sur les publics qui sont souvent un objectif principal pour les financeurs.

La LOLF représente une illustration parfaite de la détection des deux supports d'exploitation susmentionnés. Le premier support « production artistique » est représenté par le Programme 131 « Création » dont les données pour mesurer les objectifs 1, 2 et 4 sont facilement détectables. Le support polémique « Objectifs des productions », correspond à l'objectif 3 du programme 131 et les deux objectifs du Programme 224 « Transmission des savoirs et démocratisation de la culture ».

07 indicateurs (05 d'impact et 02 d'efficience) inclus dans le programme 131 concernent directement les institutions culturelles publiques. Deux autres indicateurs d'impact inclus dans le programme transversal 224 peuvent concerner ces institutions.

**Parcours des indicateurs de performance et fixation des cibles selon la LOLF**
A l'origine de la conception des indicateurs de performance d'après la LOLF dans le domaine culturel en général, un groupe de travail composé de représentants des services déconcentrés du ministère de la Culture (Directions Régionales des Affaires Culturelles « DRAC ») et des représentants des services centraux du même ministère.

Approuvés par la Direction du budget du ministère du Budget, des Comptes publics et de la Fonction publique, ces indicateurs ont été intégrés dans le premier Projet Annuel de Performance (PAP) du secteur culturel en 2006.

Chaque année, la Direction du Budget émit une circulaire pour annoncer le lancement de la procédure budgétaire et la tenue de la Conférence de performance (des réunions techniques programmées généralement vers le mois de mars). Chaque ministère a sa propre réunion qui dure une demi-journée dans laquelle le PAP, contenant les cibles (objectifs chiffrés) des indicateurs à atteindre pour l'année en cours est présenté et l'intervention sur les indicateurs par suppression, rajout ou modification est discutée, ainsi que le PAP (analyse de l'exécution de l'année précédente). Avant cette réunion, le contenu des indicateurs est discuté en interne au ministère de la Culture. De la sorte, lors de la réunion annuelle des conseillers sectoriels DRAC (Conseillers théâtre pour notre cas), organisée par la le ministère de la Culture via la Direction de la Musique, de la Danse, du Théâtre et des Spectacles (DMDTS), cette question peut être inscrite à l'ordre du jour.

La partie PAP à présenter au cours de la Conférence de performance consacré au programme 131 « Création » est rédigée par le Contrôleur de gestion du programme de la DMDTS.

La réunion regroupe en plus des membres du Bureau du budget: les représentants du programme « Création » à savoir le Responsable de programme, le Secrétaire général de programme et le Contrôleur de gestion du programme ; le Secrétariat général du ministère de la Culture représenté par le Secrétaire général du ministère, le Sous-directeur des affaires financières ainsi que le Responsable de la Mission ministérielle contrôle de gestion.

A la fin de la Conférence de performance, la Mission contrôle de gestion réalise un compte rendu sur l'ensemble de la mission culture qui lui permettra de s'assurer de la mise en œuvre des conclusions de cette réunion auprès des représentants des programmes et de diffuser l'information auprès des réseaux des contrôleurs de gestion des DRAC. Ce compte rendu aide aussi le Contrôleur de gestion du programme à réaliser les Fiches indicateurs qui présentent en détails les indicateurs de performance. Ces fiches sont retournées à la Mission ministérielle pour approbation avant d'être intégrées dans un logiciel nommé OPUS. Ce dernier, outil de collecte et de calcul des informations, permet à l'ensemble des Contrôleurs de gestion des 26 DRAC à travers le pays de visualiser les Fiches indicateurs pour organiser, selon les normes, la collecte des données.

En outre, et selon les conclusions de la Conférence de performance, le contrôleur de gestion du programme rédige le RAP de l'année n-1 ainsi que le PAP pour l'année n+1.

Les deux documents sont envoyés à la Mission ministérielle du contrôle de gestion qui les intègre dans une application interministérielle après avoir effectué une relecture et une synthèse générale des programmes qui constituent la mission « Culture ». Le RAP, annexé au Projet de Loi de Règlement (PLR), permet aux parlementaires, principaux destinataires du document, de savoir si les objectifs ont été atteints ou pas. Le PAP, annexés au Projet de Loi de Finance (PLF), leur permet de constater les objectifs fixés pour l'année à venir.

**Schéma 1 – Parcours des indicateurs performance selon la LOLF « Mission Culture »**

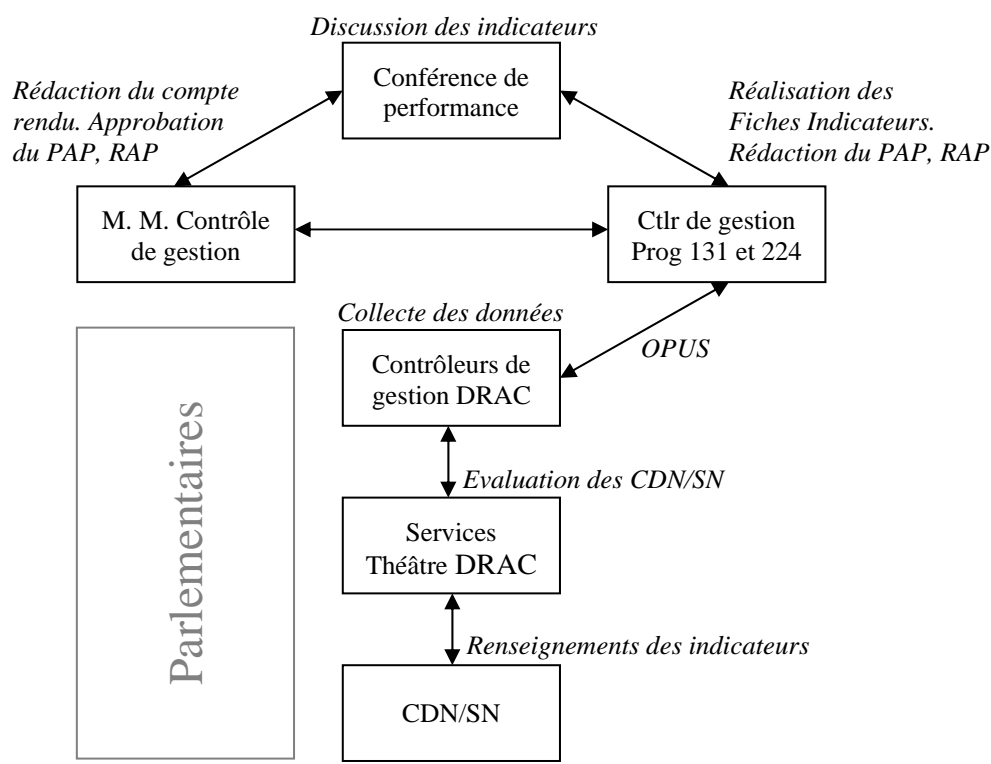

Pour la fixation des cibles-objectifs à atteindre, le Service Théâtre de la DRAC analyse l'évolution des indicateurs de performance en fonction de la Directive Nationale d'Orientation (DNO) du ministère de la Culture ainsi que la stratégie générale établie par le responsable du programme 131 (DMDTS). Il fixe ensuite les cibles en prenant en considération les données envoyées par les CDN et les SN. C'est à partir de ces cibles que le Service Théâtre évalue la performance des institutions implantées dans son territoire de compétence. Il transmet par la suite ces cibles au Contrôleur de gestion DRAC qui vérifie si la méthodologie de calcul à été respecté. En cas d'anomalie, il rencontre le responsable du Service Théâtre pour essayer de détecter la source du problème. Si cette tâche n'est pas atteinte, l'un ou l'autre contacte le Bureau de l'observation du spectacle vivant au DMDTS ou la délégation théâtre à la DMDTS chargée des CDN ou SN pour comparer ses données avec celles que reçoit ce délégué. En effet, les structures culturelles envoient leurs données et à la DRAC et à la DMDTS dans la même période (entre mai et juillet) à travers un document nommé UNIDO et un deuxième sous forme d'une enquête nationale annuelle. Les structures (directeurs et administrateurs) et la DRAC (Conseiller Théâtre) se réunissent en moyenne deux fois par an dans des réunions appelées Comités de suivi (vers le deuxième trimestre de l'année et le quatrième trimestre) ce qui leur donne l'occasion de discuter des indicateurs et des cibles. Les structures se réunissent par contre une seule fois par an avec la DMDTS (Délégation théâtre).
Le contrôleur de gestion de la DMDTS envoie lors de la préparation des documents budgétaires à l'ensemble des contrôleurs de gestion DRAC une cible nationale qu'il conçoit en s'appuyant sur un historique de trois années. Les contrôleurs de gestion DRAC doivent se positionner par rapport à cette moyenne puis justifier l'écart. Il leur est demandé par la suite d'identifier les leviers d'action pour atteindre ces cibles, c'est-à-dire de définir les moyens pour améliorer le résultat. La cible nationale à intégrer dans le PAP est réajustée selon l'appréciation du Responsable programme.
Rédigé par le contrôleur de gestion de la DMDTS et relu par la Mission ministérielle contrôle de gestion, le PAP est présenté au Bureau du budget qui commente les indicateurs et donne son aval ou pas pour les cibles proposées.

**De la polémique sur les indicateurs de performance**
Les discussions sur l'utilisation d'indicateurs quantitatifs dans l'analyse des propositions relatives à l'activité artistique et culturelle ont toujours été marquées par une tension palpable. Cependant, la théorie traduit un certain consensus que la pratique ne reconnaît pas [J.M. Schuster, 1996].

En France, contrairement à l'évaluation, pratique plus au moins acceptée par les acteurs culturels du spectacle vivant au nom du respect de la déclaration des droits de l'Homme de 1789 (*Art. 15 : La Société a le droit de demander compte à tout Agent public de son administration*), l'instauration d'une culture du résultat à travers des indicateurs de performance quantitatifs se heurte à une hostilité farouche. Pour faire valoir leur point de vue, les réfractaires à la nouvelle logique n'hésitent pas à mettre en avant le caractère sacro-saint de l'art pour lequel toute forme de quantification relèverait, selon eux, du totalitarisme. L'on peut citer par exemple la réaction d'un directeur d'un Centre dramatique nationale : « *L'objet même de la demande de vos services pose problème : en effet, à partir du tableau statistique élaboré par la direction de la musique, de la danse, du théâtre et des spectacles – DMDTS – sur l'activité déployée par la structure dans la saison passée, tableau qu'il nous faut renseigner chaque fin de saison, il nous a été demandé des extrapolations sur des prévisions 2008 (saison2007/2008) et une cible 2010 (saison 2009/2010). […] Or la construction de ce tableau ne se prête vraiment pas à une prospective à 3 ans, la nature artistique de notre activité rendant celle-ci fortement dépendante des subventions d'une part, et, pour des structures de création comme les CDN, des tournées, d'autre part. Or celles-ci sont tout à fait imprévisibles. La cible 2010 ainsi définie serait totalement arbitraire. […] Il nous paraît difficile, voire dangereux, de produire des données très incertaines dont, sans connaître l'usage qui pourra en être fait, nous pouvons penser qu'elles serviront d'une manière ou d'une autre de référence : il n'est pas innocent de transformer un outil statistique en outil de gestion. […] Aussi avons-nous décidé […] de ne pas renseigner ce tableau cible 2010, et de vous proposer de travailler avec vos services sur des indicateurs et une méthodologie plus assurée, à mettre en œuvre pour l'année prochaine* » [Y. Gaillard, 2008].

En réalité, le refus des indicateurs quantitatifs par les praticiens de la culture prend sa source dans une dialectique politique et économique beaucoup plus large que celle produite à travers les arguments qui sont utilisés de nos jours. En effet, le domaine théâtral et celui de l'action culturelle, très imprégnés par l'idéologie communiste, et plus généralement de gauche, résonnent encore selon l'équation : économie = capitalisme = droite = adversaire [X. Dupuis, 2004].
Plus loin encore, en plaçant des indicateurs de performance de démocratisation culturelle à l'intérieur du programme « Création » et en créant un programme spécifique « Transmission des savoirs et démocratisation de la culture », l'Etat met la question de l'accès à la culture pour tous au centre de la politique culturelle française, et ceci au détriment de la création. L'Etat, estimant que l'offre culturelle ne profite qu'à un public privilégié, veut développer désormais la demande culturelle et non plus l'offre culturelle.
Pour la plupart des acteurs du spectacle vivant, la démocratisation de la culture n'est qu'un leurre, pire, c'est une « tentative de standardisation de la culture » étant donné qu'il s'agit d'imposer à tout prix, pour ceux qui ne veulent pas accéder à la culture « officielle », un ensemble de formes artistiques qu'ils ne reconnaissent ou ne connaissent pas. Ainsi, pour les populations issues de la diversité par exemple, on avance l'argument de l'incompatibilité de la démocratisation de culture qui veut imposer une forme culturelle standardisée avec la Charte de l'UNESCO pour la diversité culturelle en milieu métropolitain.
Pour l'autre camp, composé surtout de décideurs politiques de droite et de techniciens de la culture (même de gauche), si la démocratisation de la culture n'a pas fonctionné, c'est parce qu'elle a été placée comme objectif second et est passée derrière la création. La LOLF, par ses indicateurs de performance, est alors venue à la rescousse d'une situation des plus critiques. Cette tendance s'est illustrée dans la lettre du Président de la république Nicolas Sarkozy adressée à la ministre de la Culture « *Les acquis de cette politique [politique culturelle des années 1980] sont considérables : une offre artistique foisonnante, des musées et des monuments rénovés, un cinéma rivalisant avec la production internationale. Ces succès ne doivent cependant pas faire oublier les lacunes et les ratés : un déséquilibre persistant entre Paris et les régions, une politique d'addition de guichets et de projets* *au détriment de la cohérence d'ensemble, une prise en compte insuffisante des publics, et surtout l'échec de l'objectif de démocratisation culturelle. De fait, notre politique culturelle est l'une des moins redistributives de notre pays. Financée par l'argent de tous, elle ne bénéficie qu'à un tout petit nombre* » [N. Sarkozy, 2007].

Sur le plan économique, le différent entre les deux tendances peut s'expliquer comme suit : de nos jours, la notion de performance prend dans son domaine de prédilection, celui des affaires, une dimension exclusivement financière. Or, la performance, dans sa fonction de rémunération du risque, se doit de désigner celui qui doit supporter ce risque. L'actualité économique, et contrairement à ce que prétend la théorie classique, démontre que c'est le salarié qui est désigné par la performance pour supporter un risque de plus ne plus grandissant, et non pas l'actionnaire (délocalisation malgré les profits gigantesques des multinationales…). Ainsi, quand la LOLF fixe par exemple comme indicateur « la recette moyenne par place offerte » ou « la part des charges fixes dans le budget des structures », cela peut s'apparenter à une volonté Etatique de transplanter des indicateurs créés à l'intérieur même du monde des affaires qui spécifie l'employé comme porteur du risque disions-nous . Dans le secteur bancaire par exemple, les banquiers utilisent majoritairement les critères comptables et financiers classiques pour définir les indicateurs internes de performance dont les deux plus importants « l'excèdent brut d'exploitation (EBE) » et « la capacité d'autofinancement ». Dans les CDN, on retrouve des indicateurs semblables, et ceci, au-delà de la LOLF même, puisque en 1995, l'arrêté portant contrats de décentralisation dramatique, et avec la fixation d'un taux de 20% de remplissage de la jauge par année, représentait déjà un indicateur qui s'appuyait sur les capacités d'autofinancement.

**Conclusion**

Complexe mais fiable, le schéma de l'évaluation de la performance selon la LOLF permet de recueillir des données très précises, institution par institution. Ces données, censés être utilisées par le ministère de la Culture pour évaluer la performance des institutions publiques du spectacle vivant, sont essentiellement destinées à être compilées en moyenne à présenter au parlementaires : les Rapports Annuels de Performance (RAP) et les Projets Annuels de Performance (PAP) sont devenus une fin en soi. Les institutions culturelles publiques du spectacle vivant sont de ce fait évaluées mais ne sont pas soumises à la culture du résultat selon ce que la LOLF prévoit : l'évaluation continue à être considérée comme un « système d'analyse » et non pas comme un « système de contrôle ». La nouveauté de l'expérience LOLF et la frilosité envers la performance à l'intérieur même des services de l'administration centrale expliquent sans doute cet état de fait. Ceci est à relier avec la conviction idéologique de chaque responsable.
Autre frein à l'application de la LOLF dans les institutions culturelles publiques du spectacle vivant, du moins celles de la production et de la diffusion du spectacle vivant, la mésentente potentielle entre les collectivités locales, financeur principal de ces institutions mais complètement en dehors du schéma de performance, et l'Etat quant au contenu des indicateurs et le taux des cibles.
Sur le plan juridique, un autre frein apparait : la définition juridique de la subvention en France exclue toute fixation d'objectifs pour les établissements subventionnés. La subvention est une somme d'argent qui n'a pas de contrepartie directe pour la personne publique. Or, l'instauration d'indicateurs de performance va à l'encontre de cette définition. La requalification des conventions liant l'Etat aux institutions culturelles du spectacle vivant en marché public ou en délégation de service public (DSP) écarte tout risque juridique pour l'application de la LOLF [J. Carabalona, N. Coppinger, 2007].

Deux perspectives se profilent dans l'horizon de la nouvelle loi organique : un échec puis un abandon comme c'était déjà le cas avec l'expérience de la Rationalisation des choix budgétaires (RCB), instaurée en1968 et abandonnée en 1984, cette option est peu probable compte tenu du consensus politique et la constitutionnalisation de la LOLF ; un développement à moyen terme qui fera que le montant des subventions dépendra entièrement ou en partie de la réalisation ou pas des indicateurs de performance.